\documentstyle[12pt]{article}

\begin{document}
\topmargin 0pt
\oddsidemargin 7mm
\headheight 0pt
\topskip 0mm

\addtolength{\baselineskip}{0.40\baselineskip}

\begin{center}

\vspace{36pt} {\large \bf SU(5/3) Superalgebra and Its
Representations of Fundamental Particles}

\end{center}

\vspace{36pt}

\begin{center}

Chang-Ho Kim$^a$ and Seung Kook Kim$^b$

{\it Department of Physics and Basic Science Research Institute, Seonam University, \\
Namwon, Jeonbuk 590-711, Korea}

\vspace{12pt}

Young-Jai Park$^c$

{\it Department of Physics,  Sogang University, Seoul 121-742, Korea \\ and \\
Center for Quantum Spacetime, Sogang University, Seoul 121-742, Korea}

\end{center}

\vspace{1cm}
\begin{center}
{\bf ABSTRACT}
\end{center}

We study the lowest dimensional typical and atypical representations of SU(5/3)
superalgebra as a possible unified gauge theory having a natural
SU(5) subalgebra with SU(3) extra structure, which will
be used to accommodate three generations of fundamental particles.
By using Kac-Dynkin weight techniques, we find that all known quarks
and leptons can be really accommodated in one atypical irreducible
representation of SU(5/3).

\vspace{2cm}


\vspace{12pt}

\noindent

\vspace{12pt}

\vfill
\hrule \vspace{0.5cm}
\hspace{-0.6cm}$^a$ E-mail address : chhjkim@gmail.com \\
$^b$ E-mail address : skandjh@empal.com \\
 $^c$ E-mail address : yjpark@sogang.ac.kr

\newpage
\section{Introduction}

There is no doubt about the Standard Model \cite{gla,wei} with gauge
group (or Lie algebra) $\rm{SU(3)_c \otimes SU(2)_w \otimes
U(1)_Y}$. However, the theory contains many unexplained patterns and
free parameters so one tries to find a simpler model by fitting the
data of the Standard Model into a larger symmetry.
The simplest model including one generation of fundamental particles is
a grand unified theory(GUT) based upon the Lie algebra SU(5) which
is proposed by Georgi and Glashow \cite{gg}. Even though the SU(5) GUT
model shows many promising features, it unified the Standard Model
only partly. For instance all known fundamental fermions cannot be
contained in a single irreducible representation (irrep) of SU(5).
Other models, for example SO(10) \cite{fm} or E6 \cite{grs},
containing SU(5) as a subalgebra were proposed \cite{ros}. Such GUTs also had a problem in the proton decay that was predicted
in the SU(5) model but not be confirmed by experiments.

From then particle physics has searched supersymmetric theories such
as supergravities, superstrings, {\it M} and {\it F} theories. There
have been many efforts to find suitable superalgebras which describe these supersymmetric theories \cite{kac,gl,hun}. Supersymmetric
extensions of Poincar\'{e} algebra in arbitrary dimensional
space-time were reviewed, and their representations(reps) for the supermultiplets of
all known supergravity theories were extensively searched by
Strathdee \cite{str,cre}. Moreover, {\it M} and {\it F} theories
\cite{sch} have been  also tackled from the point of view of the
general properties of the superalgebra \cite{bar}. Kac proposed possible connections to the
supersymmetric Standard Model with the reps of two exceptional superalgebras E(3/6) and E(3/8)
that have the maximal compact subgroup $\rm{SU(3) \otimes SU(2)\otimes U(1)}$ \cite{kac1}. Recently, Javis suggested polynomial deformations of Lie superalgebras and their reps, where the supermultiplets do not have any superpartners \cite{jav}.

On the other hand, a new unification in terms of the Lie
superalgebra SU(1/5) has been given \cite{dj,bat}, and Stoilova and
Jeugt showed some elegant features of the supersymmetric SU(5) GUT model
\cite{SJ}. Since the Lie superalgebra SU(1/5) has $\rm U(1) \otimes
SU(3)_c \otimes SU(2)_w \otimes U(1)_Y$ as a subalgebra, it
essentially contains the Lie algebra of the Standard Model. All the
fundamental fermions of a single generation (so all leptons and
quarks of left and right chirality and their antiparticles) are
inside one single irrep of SU(1/5). The first U(1) in the subalgebra
level could be considered as a tool of distinguishing between the
different generations with same irreps of $\rm U(1) \otimes SU(3)_c
\otimes SU(2)_w \otimes U(1)_Y$.

In this paper, we study typical and atypical representations of
SU(5/$n$) superalgebra as a possible unified gauge theory having a
natural SU(5) subalgebra structure with SU($n$) extra one, which
will be used to accommodate of fundamental particles. We show that the three generations of quarks and leptons
are in one atypical irrep of SU(5/3) superalgebra by using
Kac-Dynkin weight techniques. In Sec. 2, the mathematical
structure of the SU(5/$n$) superalgebra is given. In Sec. 3, we briefly
recapitulate the previous work \cite{SJ} of the SU(5/1) unification
in terms of Kac-Dynkin weight techniques. In Sec. 4, we investigate one irrep for the unification of three generations by using the SU(5/3) superalgebra, which will be shown the lowest one among SU(5/$n$) superalgebras. Here, by analyzing a branching pattern SU(5/1)$\otimes$ SU(2) $\subset$ SU(5/3), we explicitly show that all known three generations of fundamental
particles could be identified in just one irrep of the highest-highest weight $(0~0~0~0,~1,~0~0)=(\mathbf{136_B \oplus 136_F})$ of SU(5/3). The last section contains
conclusion.

\vspace{1cm}
\section{Mathematical Structures of SU(5/$n$) Superalgebra}

The Kac-Dynkin diagram of SU(5/$n$) Lie superalgebra is

\begin{eqnarray}
a_1~~~~~ a_2~~~~~  a_3~~~~~  a_4~~~~~ a_5~~~~~ a_6~~~~~ \cdots ~~~~~
a_{n+4}
\nonumber \\
\bigcirc \!\!-\!\!\!-\!\!\!-\!\!\bigcirc \!\!-\!\!\!-\!\!\!-\!\!
\bigcirc \!\!-\!\!\!-\!\!\!-\!\!\bigcirc \!\!-\!\!\!-\!\!\!-\!\!
\bigotimes \!\!-\!\!\!-\!\!\!-\!\!\bigcirc \!\!-\!\!\!-\!\! ~~
\cdots~~  \!\!-\!\!\!-\!\!\!-\!\!\bigcirc
\end{eqnarray}

\noindent where the set $(a_1~a_2~\cdots ~a_{n+4})$ determines the
highest-highest weight vector of an irrep of SU(5/$n$)
\cite{kac,gl, hun, Mieg}. Each weight component $a_i$ for $i\neq 5$ of
the highest-highest weight vector should be a nonnegative integer,
while $a_5$ could be any \emph{complex} number. The first four white
nodes corresponding $(a_1,~a_2,~a_3,~a_4)$ and the $n-1$ white nodes
for $(a_6,~\cdots, ~a_{n+4})$ denote SU(5) and SU($n$) bosonic
subalgebra, respectively, and the fifth node for the component $a_5$
is responsible to supersymmetric generators. The graded Cartan
matrix for SU(5/{$n$}) is

\vspace{0.5cm}
\begin{equation}
\left [
\begin{array}{rrrrrrrrrrr}
2  & -1 & 0  & 0 & 0  & ~0  & 0  & \cdots & 0\\
-1 & 2  & -1 & 0 & 0  & ~0  & 0  &  \cdots & 0\\
0  & -1 & 2  & -1 & 0 & ~0  & 0  &  \cdots & 0\\
0  & 0  & -1 & 2  & -1 & ~0 & 0  &  \cdots & 0\\
0  & 0  & 0  & -1 & 0  & ~1 & 0  &  \cdots & 0\\
0  & 0  & 0  & 0  & -1 & ~2 & -1 & \cdots & 0 \\
\cdot  & \cdot  & \cdot  & ~\cdot & \cdot  & -1 & 2 &  \cdots & 0\\
\cdot  & \cdot  & \cdot  & ~\cdot & \cdot  & \cdot & \cdot &  \ddots & -1\\
0  & 0  & 0  & 0 & 0  & \cdots & ~0 & -1 & 2\\

\end{array}
\right ]
 .
\end{equation}
\vspace{0.5cm}

\noindent Note that each positive/negative simple even root
$\alpha_i^\pm(i=1,\cdots,n+4, i\neq5)$ corresponds to the $i$-th
column of the graded Cartan matrix, while the positive/negative
simple odd root $\beta_5^{5\pm}$ corresponds to the 5-\textit{th} column of
the matrix. Other odd roots are obtained by

\begin{equation}
\beta_5^{i\pm} = [\alpha_i^\pm ,~ \beta_5 ^{(i+1) \pm} ],~~~
i=1,~2,~3,~4,
\end{equation}

\noindent and

\begin{equation}
\beta_{j+1}^{5\pm} = [\beta_j ^{5 \pm}, \alpha_j^\pm ],~~~
j=5,~6,\cdots,~n+4.
\end{equation}

In general, the irreps of SU($m$/$n$) are divided into two types,
which are \emph{typical} and \emph{atypical} \cite{kac,hun}. All
atypical irreps of SU(5/$n$) are characterized by the fifth
component $a_5$ of the highest-highest weight. The atypicality
condition of SU(5/$n$) is given by

\begin{equation}
a_5 = \sum_{k=6}^{j}a_k  -\sum_{k=i}^{4}a_k -10 + i + j,~~~1\leq
i\leq 5, ~ 5\leq j\leq n-1. \label{line1}
\end{equation}

\noindent The typical irreps of SU(5/$n$) consist of $5n+1$ floors,
and have equal bosonic and fermionic degrees of freedom. An atypical
irrep is obtained by terminating some odd root strings in a full
weight system when $a_5$ satisfies the relation in Eq. (5) for
specific \emph{i}'s and \emph{j}'s.

The lowest dimensional typical rep of SU(5/$n$) is
$(0~0~0~0,~a_5,~0~\cdots~0)$ = $2^{5n-1}_B~ \oplus ~2^{5n-1}_F$, for
$a_5\neq n-1, \cdots,2,1,0,-1,-2,-3,-4$, where the subscripts $B$
and $F$ denote bosons and fermions respectively, such as

\begin{equation}
\begin{array}{lc}
\\
\mid \mbox{gnd}>~~~~~~~~~~~~~~~~~~~~~~~~~~~~~~~~~~(0~0~0~0,~a_5,~0~\cdots~0) \\
 ~~~~~~~~~~~~~~~~~~~~~~~~~~~~~~~~~~~~~~~~~~~~~~~~~~~~~~~~~~~\downarrow \beta^{5-}_5 \\
\mid \mbox{1st}>~~~~~~~~~~~~~~~~~~~~~~~~~~~~~~~~~~~~~~(0~0~0~1,~a_5,~1~0~\cdots~0) \\
~~~~~~~~~~~~~~~~~~~~~~~~~~~~~~~~~~~~~~~~~~~~~~~~\swarrow \beta^{4-}_5 ~~~~~~~~~~~ \searrow \beta^{5-}_6\\
\mid \mbox{2nd}>~~~~~~~~~~~~~~(0~0~1~0,~a_5+1,~2~0~\cdots~0)~~~(0~0~0~2,~a_5-1,~0~1~\cdots~0)\\
~~~~~~~~~~~~~~~~~~~~~~~~~~~~\swarrow \beta^{3-}_5 ~~~~~~~~~ \searrow \beta^{5-}_6~~~~~~~~~~~\swarrow \beta^{4-}_5 ~~~~~~~~ \searrow \beta^{5-}_7 \\
\mid \mbox{3rd}>  (0~1~0~0,~a_5+2,~3~0~\cdots~0)~(0~0~1~1,~a_5,~1~1~0~\cdots~0)~(0~0~0~3,~a_5-2,~0~0~1~\cdots~0)\\
~~~~~~~~~~~~~~~~~~~~~~~~~~~~~~~~~~~~~~~~~~~~~~~~~~~~~~~~~~~~\vdots \\
~~~~~~~~~~~~~~~~~~~~~~~~~~~~~~~~~~~~~~~~~~~~~~~~~~~~~~~~~~~~\vdots \\
\mid \mbox{(5$n$-1)-th}>~~~~~~~~~~~~~~~~~~~~~~~~(1~0~0~0,~a_5-n+5,~0~\cdots~0~1)\\
~~~~~~~~~~~~~~~~~~~~~~~~~~~~~~~~~~~~~~~~~~~~~~~~~~~~~~~~~~~\downarrow \beta^{1-}_{n+4} \\
\mid \mbox{5$n$-th}>
~~~~~~~~~~~~~~~~~~~~~~~~~~~~~~(0~0~0~0,~a_5-n+5,~0~\cdots~0)
\end{array}
\end{equation}

\noindent The even and odd floors alternate between fermions and
bosons by the successive actions of odd roots.

If one take the value $a_5$ to satisfy the atypical condition, then
the corresponding odd root strings are terminated from the typical
irrep in Eq. (6), hence the the weight system
$(0~0~0~0,~a_5,~0~\cdots~0)$ should be separated in several atypical
irreps.

The fundamental rep of SU(5/$n$) is $(1~0~\cdots~0)$ which has the
substructure of $(\mathbf{5 \oplus n})$ in SU(5)$\otimes$ SU($n$)
$\otimes$ U(1) bosonic subalgebra basis as follows:
\\
\begin{equation}
\begin{array}{lccl}
\mid \mbox{gnd}> & (1~0~0~0,~0,~0~0\cdots~0) & = & \mathbf{(5,1)}_n\\[6pt]
\mid \mbox{1st}> & (0~0~0~0,~1,~1~0\cdots~0) & = & \mathbf{(1,n)}_5 , \\
\end{array}
\end{equation}\\

\noindent where the subscripts denote U(1) values corresponding the
weight component $a_5$. The U(1) supercharge generator is
$Diag(n,n,n,n,n,\underbrace{5,\cdots,5}_{n})$ to satisfy the
supertraceless condition.

The complex conjugation of the fundamental rep of SU(5/$n$) for
$n\geq2$ is $(0~\cdots~0~1)=(\mathbf{n^* \oplus 5^*})$ such as

\begin{equation}
\begin{array}{lccl}
\mid \mbox{gnd}> & (0~0~0~0,~0,~0~\cdots~0~1) & = & \mathbf{(1,n^*)}_{-5} \\[6pt]
\mid \mbox{1st}> & (0~0~0~1,-1,~0~\cdots~0~0) & = & \mathbf{(5^*,1)}_{-n}~. \\
\end{array}
\end{equation}
\\

\noindent The Eq. (8) reduces to $(0~0~0~0,~-1)$ for the case
SU(5/1):

\begin{equation}
\begin{array}{lccl}
\mid \mbox{gnd}> & (0~0~0~0,-1) & = & \mathbf{1}_{-5} \\[6pt]
 \mid \mbox{1st}> & (0~0~0~1,-1) & = & \mathbf{5^*}_{-1}. \\
\end{array}
\end{equation}
\\

Like an usual algebra SU($5+n$) the superalgebra SU(5/$n$), where $n\neq5$, has
$(5+n)^2-1$ roots which consists of even roots for the subalgebras SU(5), SU($n$) and U(1) plus $10n$ odd roots. The even and odd roots are in the adjoint rep $(1~0~\cdots~0~1)$ such as
\\
\begin{equation}
\begin{array}{llll}
\mid \mbox{gnd}> & (1~0~0~0,~0,~0\cdots~0~1) & = & \beta_i^{j+}\\[6pt]
\mid \mbox{1st}> & (1~0~0~1,-1,~0~\cdots~0) & = & \mbox{SU(5)} \\
                 & (0~0~0~0,~0,~0~\cdots~0) & = & \mbox{U(1)} \\
                 & (0~0~0~0,~1,~1~0~\cdots~0~1) & = & \mbox{SU($n$)}\\[6pt]
\mid \mbox{2nd}> & (0~0~0~1,~1,~1~0~\cdots~0) & = & \beta_i^{j-}. \\
\end{array}
\end{equation}
\\

\noindent Note that Eq. (10) becomes $(1~0~0~0,~-1)$ and $(1~0~0~0,~0,~1)$
for SU(5/1) and SU(5/2), respectively as following

\begin{equation}
\begin{array}{llll}
\mid \mbox{gnd}> & (1~0~0~0,-1) & = & \beta_5^{j+}\\[6pt]
\mid \mbox{1st}> & (1~0~0~1,-1) & = & \mbox{SU(5)} \\
                 & (0~0~0~0,~0) & = & \mbox{U(1)}\\[6pt]
\mid \mbox{2nd}> & (0~0~0~1,~0) & = & \beta_5^{j-}. \\
\end{array}
\end{equation}
\\

\begin{equation}
\begin{array}{llll}
\mid \mbox{gnd}> & (1~0~0~0,~0,~1) & = & \beta_i^{j+}\\[6pt]
\mid \mbox{1st}> & (1~0~0~1,-1,~0) & = & \mbox{SU(5)} \\
                 & (0~0~0~0,~0,~0) & = & \mbox{U(1)} \\
                 & (0~0~0~0,~1,~2) & = & \mbox{SU(2)}\\[6pt]
\mid \mbox{2nd}> & (0~0~0~1,~1,~1) & = & \beta_i^{j-}, \\
\end{array}
\end{equation}

\vspace{1cm}
\section{Typical and Atypical Representations of SU(5/1)}

The lowest dimensional typical irrep of SU(5/1) is $(0~0~0~0,~a_5)$ with $a_5\neq
-4,-3,-2,-1,0$ such as

\begin{equation}
\begin{array}{lccc}
\mid \mbox{gnd}> & (0~0~0~0,~a_5) & = & \mathbf{1} \\
                 & \downarrow \beta^{5-}_5 \\
\mid \mbox{1st}> & (0~0~0~1,~a_5) & = & \mathbf{5}^* \\
                 & \downarrow \beta^{4-}_5 \\
\mid \mbox{2nd}> & (0~0~1~0,~a_5+1) & = & \mathbf{10}^* \\
                 & \downarrow \beta^{3-}_5 \\
\mid \mbox{3rd}> & (0~1~0~0,~a_5+2) & = & \mathbf{10} \\
                 & \downarrow \beta^{2-}_5 \\
\mid \mbox{4th}> & (1~0~0~0,~a_5+3) & = & \mathbf{5}\\
                 & \downarrow \beta^{1-}_5 \\
\mid \mbox{5th}> & (0~0~0~0,~a_5+4) & = & \mathbf{1}.\\
\end{array}
\end{equation}
\\

\noindent From the weight system $(0~0~0~0,~a_5)$ one can make
singlet $(0~0~0~0,~0)$ by taking $a_5=0$. In this case a
positive/negative odd root $\beta^{5\pm}_5$ must be destroyed in
Eq. (13). Notice that the other five floors make another atypical rep
$(0~0~0~1,~0)$. Indeed, the Eq. (13) is separated into two independent atypical irreps by
removing one of the positive/negative odd roots $\beta^{i\pm}_5$'s by using the atypical condition
in Eq. (5). We get two independent
atypical reps, where the first one is starting from the ground floor
and another one composed by remaining floors such as

\begin{equation}
\begin{array}{llll}
a_5=0:  &(0~0~0~0,~~0)~&\oplus~&(0~0~0~1,~0) \\
a_5=-1: &(0~0~0~0,-1)~&\oplus~&(0~0~1~0,~0) \\
a_5=-2: &(0~0~0~0,-2)~&\oplus~&(0~1~0~0,~0) \\
a_5=-3: &(0~0~0~0,-3)~&\oplus~&(1~0~0~0,~0) \\
a_5=-4: &(0~0~0~0,-4)~&\oplus~&(0~0~0~0,~0).\\
\end{array}
\end{equation}
\\

Recently, Stoilova and Jeugt had shown that the fundamental fermions
fit inside just one irrep of SU(1/5)\cite{SJ}. Here we briefly
reconstruct the theory in terms of SU(5/1) superalgebra.

The typical irrep $(0~0~0~0,~a_5)$ has substructures that are
relevant to the supersymmetric extension of SU(5) model. Since the
rep contains fermionic and bosonic parts simultaneously, the odd
floors $[\mathbf{5}^* \oplus \mathbf{10} \oplus \mathbf{1}]$ contain the original fermions, while the even floors $[\mathbf{1}
\oplus \mathbf{10}^* \oplus \mathbf{5}]$ could be identified with the superpartners.

We obtain the branching rule such as

\begin{equation}
\begin{array}{llll}
          & ~~~\textrm{SU}(5/1) & \textrm{SU}(5) \otimes \textrm{U}(1) &  \textrm{SU}_c (3) \otimes  \textrm{SU}_w (2) \otimes  \textrm{U}_Y (1)\\
\hline
\mid \mbox{gnd}>        & (0~0~0~0,~a_5)~~~~& (0~0~0~0)_{5a_5}~~~~~&~~~~~~(0~0)(0)_0\\[6pt]
\mid \mbox{1st}>           & (0~0~0~1,~a_5)~~~~& (0~0~0~1)_{5a_5+4} &~~~~~~(0~0)(1)_{1/2}\\
                           &                &                       &~~~~~~(0~1)(0)_{-1/3}\\[6pt]
\mid \mbox{2nd}>           & (0~0~1~0,~a_5+1) & (0~0~1~0)_{5a_5+8} &~~~~~~(0~0)(0)_1\\
                           &                  &                    &~~~~~~(0~1)(1)_{1/6}\\
                           &                  &                    &~~~~~~(1~0)(0)_{-2/3}\\[6pt]
\mid \mbox{3rd}>           & (0~1~0~0,~a_5+2) & (0~1~0~0)_{5a_5+12} &~~~~~~(0~1)(0)_{2/3}\\
                           &                  &                     &~~~~~~(1~0)(1)_{-1/6}\\
                           &                  &                     &~~~~~~(0~0)(0)_{-1}\\[6pt]
\mid \mbox{4th}>           & (1~0~0~0,~a_5+3) & (1~0~0~0)_{5a_5+16} &~~~~~~(1~0)(0)_{1/3}\\
                           &                &                       &~~~~~~(0~0)(1)_{-1/2}\\[6pt]
\mid \mbox{5th}>           & (0~0~0~0,~a_5+4) & (0~0~0~0)_{5a_5+20} &~~~~~~(0~0)(0)_0\\
\end{array}
\end{equation}
\\

\noindent Note that an atypical rep $(0~0~0~0,~-4)$ of SU(5/1)
exactly accommodates well known one generation $[\mathbf{5^* \oplus
10}]$ of SU(5) by decoupling the singlet fermion at the fifth floor
in Eq. (13) which is included in the SO(10) model, while
$(0~0~0~1,~0)$ which is complex conjugation of $(0~0~0~0,~-4)$
contain $[\mathbf{5^* \oplus 10 \oplus 1}]$ as fermionic states.
Another atypical rep $(0~0~0~0,~-3)$ also contain $[\mathbf{5^*
\oplus 10}]$ but ends at the third floor.

We find an atypical rep $(0~0~0~0,~1,~0)=(\mathbf{56_B \oplus
56_F})$ of SU(5/2) contains $(0~0~0~0,~1)$, which is typical, and
$(0~0~0~1,~0)$, which is atypical irrep of SU(5/1) simultaneously, such as

\begin{equation}
\begin{array}{lcl}
(0~0~0~0,~1,~0) &\rightarrow &(0~0~0~0,~1)_5 \oplus (0~0~0~1,~0)_8
\oplus (0~0~1~0,~0)_{11}\\
&&\oplus (0~1~0~0,~0)_{14} \oplus (1~0~0~0,~0)_{17} \oplus
(0~0~0~0,~0)_{20}.
\end{array}
\end{equation}
\\

As gauge bosons are identified with the adjoint rep $(1~0~0~1)$ of SU(5) model,
we introduce the adjoint rep $(1~0~0~0,~-1)$ of SU(5/1)\cite{SJ}.
The branching rule is

\begin{equation}
\begin{array}{llll}
          & ~~\textrm{SU}(5/1) & \textrm{SU}(5) \otimes \textrm{U}(1) &  \textrm{SU}_c (3) \otimes  \textrm{SU}_w (2) \otimes  \textrm{U}_Y (1)\\
\hline
\mid \mbox{gnd}> & (1~0~0~0,-1) &~~(1~0~0~0)_{-1} & ~~~~~~~(1~0)(0)_{1/3}\\
                    &              &               & ~~~~~~~(0~0)(1)_{-1/2}\\[6pt]
\mid \mbox{1st}> & (1~0~0~1,-1) &~~(1~0~0~1)_0 & ~~~~~~~(1~0)(1)_{5/6}\\
                 &              &              & ~~~~~~~(1~1)(0)_0\\
                 &              &              & ~~~~~~~(0~0)(2)_0\\
                 &              &              & ~~~~~~~(0~0)(0)_0\\
                 &              &              & ~~~~~~~(0~1)(1)_{-5/6}\\[6pt]
                 & (0~0~0~0,~0) &~~(0~0~0~0)_0 & ~~~~~~~(0~0)(0)_0\\[6pt]
\mid \mbox{2nd}> & (0~0~0~1,~0) &~~(0~0~0~1)_1 & ~~~~~~~(0~0)(1)_{1/2}\\
                 &             &               & ~~~~~~~(0~1)(0)_{-1/3}\\
\end{array}
\end{equation}
\\

\noindent Indeed, the first floor of Eq. (17) contains the adjoint rep $(1~0~0~1)=\textbf{24}$ of SU(5) 
which represents the gauge bosons. 

\vspace{1cm}
\section{SU(5/3) Unification}

SU(5/3) has fifteen positive/negative odd roots such that an typical irrep of
the superalgebra has sixteen floors. The lowest dimensional typical irrep of SU(5/3) is
$(0~0~0~0,~a_5,~0~0)=(\mathbf{2^{14}_B \oplus 2^{14}_F})$ with
$a_5\neq -4,-3,-2,-1,0,1,2$.

Since the SU(5/3) superalgebra has SU(5/1)$\otimes$SU(2) as a subalgebra,
one may expect to find one irrep of SU(5/3) which can accommodate
three generations of fundamental particles. In particular we will
concentrate on the SU(5/1)$\otimes$SU(2) branching from SU(5/3) because both the irreps
$(0~0~0~0,~1)$ and $(0~0~0~1,~0)$ of SU(5/1) superalgebra
can fit $[\mathbf{5^*+10}]$ reps of SU(5) GUT separately. As a natural extension of Eq. (16), we note on an
atypical irrep $(0~0~0~0,~1,~0~0)=(\mathbf{136_B \oplus 136_F})$ of SU(5/3) which has the explicit weight system as follows

\begin{equation}
\begin{array}{llll}
\mid \mbox{gnd}> & (0~0~0~0,~1,~0~0)\\[6pt]
\mid \mbox{1st}> & (0~0~0~1,~1,~1~0)\\[6pt]
\mid \mbox{2nd}> & (0~0~1~0,~2,~2~0)\\[6pt]
\mid \mbox{3rd}> & (0~1~0~0,~3,~3~0)\\[6pt]
\mid \mbox{4th}> & (1~0~0~0,~4,~4~0)\\[6pt]
\mid \mbox{5th}> & (0~0~0~0,~5,~5~0).\\
\end{array}
\end{equation}
\\

\noindent The branching rule SU(5/3)$\rightarrow$SU(5/1)$\otimes$SU(2) of Eq. (18) is given by

\begin{equation}
\begin{array}{lcl}
(0~0~0~0,~1,~0~0) &\rightarrow &(0~0~0~0,~1)(0) \oplus (0~0~0~1,~0)(1) \oplus (0~0~1~0,~0)(2)\\
&&\oplus (0~1~0~0,~0)(3) \oplus (1~0~0~0,~0)(4) \oplus (0~0~0~0,~0)(5).
\end{array}
\end{equation}
\\

\noindent Note that the irrep $(0~0~0~0,~1,~0~0)$ contain one typical irrep $(0~0~0~0,~1)(0)$ and one atypical irrep $(0~0~0~1,~0)(1)$ at SU(5/1)$\otimes$SU(2) stage. If we break the SU(2) flavor part, then it can be easily shown that the irrep $(0~0~0~0,~1,~0~0)$ contains three copies of $(\mathbf{5}^* \oplus \mathbf{10})$ reps of SU(5) GUT.
\\

On the other hand, the branching rule SU(5/3)$\rightarrow$SU(5/1)$\otimes$SU(2) for the adjoint rep $(1~0~0~0,~0,~0~1)$ is given by

\begin{equation}
(1~0~0~0,-1)(0) \oplus (1~0~0~0,~0)(1)
\oplus (0~0~0~0,-1)(1) \oplus (0~0~0~0,~0)(2) \oplus (0~0~0~0,~0)(0).
\end{equation}

\noindent Similar to the SU(5/1) case in Eq. (17), the first floor of $(1~0~0~0,-1)(0)$ in Eq. (20) also contains the adjoint rep $(1~0~0~1)=\textbf{24}$ of SU(5).

\vspace{1cm}
\section{Conclusion}

In this paper we have studied a group theoretical framework for
unified model building including three generations of fundamental particles in SU(5/3). Elementary
particles is not our field of specialization, so we leave it to the
specialists to consider this representation theoretic picture as a
basis for real models. It is only after such considerations that the
proposed SU(5/3) superalgebra, which has SU(5/1)$\otimes$SU(2) substructure, can be regarded as an interesting part of
physics, or whether it is just a mathematical coincidence. In fact the SU(5/3) superalgebra could describe supersymmetric SU(5/1) unification, which includes the well-known SU(5) model [3], while the SU(2) flavor part accommodates all known three generations.

In conclusion, we have studied the lowest dimensional typical irrep and its atypical irreps of SU(5/3) superalgebra as a possible unified gauge theory that has a natural SU(5/1)$\otimes$SU(2) subalgebra structure.
As a result, we have explicitly shown that all known three
generations of fundamental particles could be in just one irrep $(0~0~0~0,~1,~0~0)$ of SU(5/3).

\vspace{1cm}

\begin{center}
{\bf Acknowledgments}
\end{center}

Y.-J. Park was supported by the National Research Foundation of Korea (NRF) grant
funded by the Korea government (MEST) through the Center for Quantum
Spacetime (CQUeST) of Sogang University with grant number
2005-0049409.

\end{document}